\documentclass[reprint,fleqn,11pt]{revtex4}
\usepackage{mathrsfs,graphicx,amsmath,amssymb}
\usepackage{hyperref}\hypersetup{colorlinks=true,linkcolor=blue,citecolor=blue,filecolor=blue,urlcolor=blue,linktocpage}

\newcommand\lae[1]{\label{#1}}

\def\lt{\left}
\def\rt{\right}
\def\f{\frac}
\def\bq{\begin{equation}}
\def\ee{\end{equation}}
\def\md{\,\mbox{d}}
\def\bs{\boldsymbol}
\def\bk{k_{\mbox{\tiny B}}}
\def\de{\delta}

\def\q{\quad}
\def\bC{\pmb{\mathscr{C}}}
\def\C{\mathscr{C}}

\def\wt{\widetilde}

\newcommand\tn[1]{\bs{\mathsf{#1}}}

\begin{document}
\baselineskip 18pt

\title{Transport coefficients of multi-component mixtures of noble gases based on {\it ab initio} potentials. Viscosity and thermal conductivity.}

\author{Felix Sharipov}
\email{sharipov@fisica.ufpr.br}
\homepage{http://fisica.ufpr.br/sharipov}
\affiliation{Departamento de F\'\i sica, Universidade Federal do Paran\'a, Curitiba, 81531-990, Brazil}

\author{Victor J. Benites}
\email{vjben@fisica.ufpr.br}
\affiliation{Universidade Positivo, Curitiba, 81280-330, Brazil}

\begin{abstract}

The viscosity and thermal conductivity of binary, ternary and quaternary mixtures of helium, neon, argon, and krypton at low density are computed for wide ranges of temperature and molar fractions, applying the Chapman-Enskog method. {\it Ab initio} interatomic potentials are employed in order to calculate the omega-integrals. The relative numerical errors of the viscosity and thermal conductivity do not exceed $10^{-6}$ and 10$^{-5}$, respectively. The relative uncertainty related to the interatomic potential is about 0.1\%. A comparison of the present data with results reported in other papers available in the literature shows a significant improvement of accuracy of the transport coefficients considered here.


\end{abstract}

\maketitle

\section{Introduction}

The technique to calculate viscosity and thermal conductivity for binary gaseous  mixtures is well elaborated and published in numerous papers, see e.g., Refs. \cite{Tip02,Tip03,Tom04,Tom05,Son49,Sha110,Sha118,Sha126}. The approach of these works is based on the Chapman-Enskog method  \cite{Cha04,Fer02} applied to a system of the kinetic Boltzmann equations. This method and experimental data on the transport coefficients were analyzed by Kestin {\it et al.}  \cite{Kes01} in order to derive empirical expressions of viscosity and thermal conductivity  for all kinds of mixtures of the  noble gases.

In practice, one deals with ternary and quaternary mixtures of noble gases as often as with binary ones, see e.g. Refs. \cite{Loy36,Ben03,Sza06,Yak02}, so that reliable data on the transport coefficients for ternary, quaternary, etc. mixtures are also of practical and scientific interest. These coefficients are included in the Navier-Stokes equations describing fluid flows in continuous medium regime. Moreover, the viscosity coefficient is important in rarefied gas dynamics \cite{Sha02B} in order to determine the equivalent-free-path used to calculate the rarefaction parameter \cite{Fre04,Sha56,Sha100,HoM03,Val10,Gos03} velocity slip  \cite{Sha54,Sha84}, and temperature jump  \cite{Sha55} coefficients. The rarefaction parameter, velocity slip, and temperature jump are widely used in modelling of micro and nano flows of gases.

The general theory of the transport coefficients described by Ferziger \& Kaper  \cite{Fer02} is valid for an arbitrary number of gaseous species. The final expressions of the coefficients are cumbersome and given in term of solution of a large system of algebraic equations. The matrix elements of the system and free terms are linear combinations of multi-fold integrals which depend on the interatomic potential. To overcome the great numerical difficulties, some approximate formulas of the transport coefficients were proposed, see e.g. the papers  \cite{Bro05,Sin02,Avs01}, which contain some fitting parameters, usually, extracted from experimental data.

Nowadays, {\it ab initio} potentials for all pairs that can be composed from helium, neon, argon, and krypton are available in the open literature, see e.g. Refs. \cite{Azi03,Cyb01,Hal01,Cac01,Hel03,Pat02,Prz01,Cen01,Jag02,Jag03,Jag04}. This information allows us to obtain the transport coefficient of any multi-component mixture of these noble gases.

Some binary mixture were considered in previously published papers, namely, the helium-neon mixture was considered in Ref. \cite{Sha118}, the transport coefficients of helium-argon and neon-argon mixtures were reported in Ref. \cite{Sha126}, and the same coefficients of helium-krypton mixture were calculated in Ref. \cite{Jag04}. However, the transport coefficient of ternary and quaternary mixtures of the noble gases have not been calculated yet on the basis of {\it ab initio} potentials. Accurate results for some binary mixtures of the noble gases are also absent in the literature.

In the present paper, numerical results on viscosity and thermal conductivity of binary, ternary and quaternary mixtures of helium, neon, argon and krypton in the temperature range from 50 K to 5000 K based on {\it ab initio} potentials are reported. The quantum approach to the interatomic interactions is employed for all kinds of collisions. The relative numerical error of the viscosity and thermal conductivity is less then $10^{-6}$ and $10^{-5}$, respectively. The estimated relative uncertainty due to the interatomic potential does not exceed 0.1\% and in some cases can be even smaller.

\section{Statement of the problem.}

Here, we consider a mixture of $K$ monatomic gases at a temperature $T$ and pressure $p$. The number density of each species is denoted as $n_i$ ($1\le i\le K$). The chemical composition of the mixture can be characterized by the mole fraction defined as
\begin{equation}
x_i=n_i/n,\q n=\sum_{i=1}^K n_i.
\lae{AX}\end{equation}
The mixture pressure is assumed to be so low that the state equation corresponds to ideal gas, i.e. $p=n\bk T$, where $\bk$ is the Boltzmann constant.

We are going to calculate the dynamic viscosity $\mu$ and thermal conductivity $\kappa$ of binary, ternary and quaternary mixtures of helium, neon, argon and krypton as a function of the temperature $T$ and chemical composition $x_i$. The calculations are based on {\it ab initio} potentials available in the open literature.

The viscosity $\mu$ is well defined in fluid mechanics, see e.g. Ref. \cite{Lan05}, while the thermal conductivity requires some clarifications. According to the papers \cite{Muc01,Fer02}, there are two kinds of thermal conductivity coefficients of mixture: partial coefficient $\kappa'$ and steady state coefficient $\kappa$. The former one $\kappa'$ characterizes a heat transfer through a mixture with an uniform chemical composition. In this case, each species of the mixture moves due to the thermal diffusion phenomenon, while the whole mixture is at rest. The latter coefficient $\kappa$ corresponds to situation when the thermal diffusion is compensated by diffusion and a time-independent mole fraction distribution is established. Under this condition, all species of the mixture are at rest. The coefficients are coupled to each other by, see Sec 6.3 from the book  \cite{Fer02},
\begin{equation}
\kappa'=\kappa +n\bk \sum_{i=1}^K k_{Ti}D_{Ti},
\lae{AV}\end{equation}
where $k_{Ti}$ is the thermal diffusion ratio of species $i$ and $D_{Ti}$ is the thermal diffusion coefficient related as
\begin{equation}
\sum_{j=1}^KD_{ij}k_{Ti}=D_{Ti},
\lae{AW}\end{equation}
with $D_{ij}$ being the multi-component diffusion coefficient. Like our previous papers, we are going to calculate the steady state thermal coefficient $\kappa$. Once the coefficients $D_{ij}$ and $D_{Ti}$ are known, the partial coefficient $\kappa'$ can be calculated too.

\section{Method of calculation.}

\subsection{Expressions of transport coefficients}

The expressions of the transport coefficients for multi-component mixture are derived by the Chapman-Enskog method applied to the kinetic Boltzmann equation in the book by Ferziger \& Kaper  \cite{Fer02}. In case of binary mixture, this method is well described in the book  \cite{Cha04} and the papers \cite{Tip02,Tip03,Tom04,Tom05}. The expression for viscosity $\mu$ of multi-component obtained in Ref. \cite{Fer02} in term of bracket integrals are used here with slightly different notations. Each bracket integral contains information about only two gaseous species  so that the expressions of these integrals obtained in Refs.  \cite{Tip02,Tip03,Tom04,Tom05} can be used here for multi-component mixture. Since the book \cite{Fer02} does not provide the explicit expression of $\kappa$, its derivations for binary mixture \cite{Tip03} is generalized for  multi-component mixture in Appendix to the present paper.

Following the previously published derivations \cite{Fer02,Tip02,Tip03,Tom04,Tom05} and those given in Appendix, the viscosity $\mu$ and thermal conductivity $\kappa$ are calculated as
\bq
\mu= \f52\bk T\sum_{i=1}^K x_i b^{(1)}_{i},
\lae{AD}\ee
\bq
\kappa=\f{75}{8}\bk^2T\sum_{i=1}^K \f{x_i}{\sqrt{m_i}}a^{(1)}_i,
\lae{QR}\ee
respectively. The coefficients $a_i^{(1)}$ and $b_i^{(1)}$ are calculated from the corresponding systems of algebraic equations
\bq
\sum_{j=1}^{K}\sum_{q=1}^N A_{ij}^{(pq)} a_j^{(q)}=\f{x_i\delta_{p1}}{\sqrt{m_i}},
\lae{AN1}\ee
\bq
\sum_{j=1}^K\sum_{q=1}^{N} B_{ij}^{(pq)} b^{(q)}_{j}=x_i\delta_{p1},\q
\lae{AN2}\ee
where $1\le i\le K$, $1\le p\le N$, $\delta_{pq}$ is the  Kronecker delta, and $N$ is the order of approximation. The values of $\mu$ and $\kappa$ converge to their exact values in the limit $N\to\infty$. The matrices $A_{ij}^{(pq)}$, $B_{ij}^{(pq)}$ are expressed in terms of the bracket integrals as
\[
A_{ii}^{(pq)}=
x_i^2\lt[S_{3/2,i}^{(p)}\bC_i,S_{3/2,i}^{(q)}\bC_i\rt]_i
\]
\begin{equation}\hskip1cm
+\sum_{\substack{j=1\\j\neq i}}^N x_ix_j\lt[S_{3/2,i}^{(p)}\bC_i,S_{3/2,i}^{(q)}\bC_i\rt]_{ij},
\lae{AO}\end{equation}
\begin{equation}
A_{ij}^{(pq)}=x_ix_j\lt[S_{3/2,i}^{(p)}\bC_i,
S_{3/2,j}^{(q)}\bC_j\rt]_{ij},\q i\neq j,
\lae{AQ}\end{equation}
\[
B_{ii}^{(pq)}=
x_i^2\lt[S_{5/2,i}^{(p-1)}\tn{C}_i,S_{5/2,i}^{(q-1)}\tn{C}_i\rt]_i
\]
\bq\hskip1cm
+\sum_{\substack{j=1\\j\neq i}}^N x_ix_j\lt[S_{5/2,i}^{(p-1)}\tn{C}_i,S_{5/2,i}^{(q-1)}\tn{C}_i\rt]_{ij}
\lae{AF}\ee
\bq
B_{ij}^{(pq)}=x_ix_j\lt[
S_{5/2,i}^{(p-1)}\tn{C}_i,
S_{5/2,j}^{(q-1)}\tn{C}_j\rt]_{ij},\q i\neq j
\lae{AI}\ee
where $S_{\nu,i}^{(p)}$ are the Sonine polynomials with the argument $\C_i^2$, i.e.
\begin{equation}
S_{\nu,i}^{(p)}
=\sum_{n=0}^p\f{\Gamma(\nu+p+1)}{(p-n)!n!\Gamma(\nu+n)}
\lt(-\C^2_i\rt)^n.
\lae{AL}\end{equation}
The tensor $\tn{C}_i$ is defined as
\begin{equation}
\tn{C}_i=\bC_i\bC_i-\f13\C^2_i \tn{I},
\lae{AG}\end{equation}
with $\tn{I}$ being the identity tensor. The dimensional molecular velocity $\bC_i$ is defined for each species as
\begin{equation}
\bC_i=\sqrt{\f{m_i}{2\bk T}}{\bs C}_i,\q  {\bs C}_i={\bs c}_i-{\bs u},
\lae{AH}\end{equation}
where $m_i$ is the atomic mass of species $i$, ${\bs c}_i$ is its molecular velocity, and ${\bs u}$ is the hydrodynamic velocity of the mixture. Some of the bracket integrals are given in the book  \cite{Fer02}. The general expressions of bracket integrals for arbitrary orders $p$ and obtained in the papers \cite{Tom04,Tom05} for binary mixtures can be used here. The first and second brackets in Eq.(\ref{AO}) are given by Eqs.(119) and (117) from  Ref. \cite{Tom04}, respectively. The brackets in Eq.(\ref{AQ}) are given by Eq.(115) from Ref. \cite{Tom05}. The first and second brackets in Eq.(\ref{AF}) are given by Eqs.(113) and (111) from  Ref. \cite{Tom05}, respectively. The brackets in Eq.(\ref{AI}) are given by Eq.(109) from Ref. \cite{Tom05}. To generalize the corresponding expressions given in the papers \cite{Tom05,Tom04} to a multi-component mixture, the subscripts ``1'' and "2" are replaced by ``i'' and ``j'', respectively. In case of $i>j$, the symmetry relations
\begin{equation}
A^{(pq)}_{ij}=A_{ji}^{(qp)},\q B^{(pq)}_{ij}=B_{ji}^{(qp)}
\lae{AS}\end{equation}
are employed.

\subsection{Transport cross sections}

The matrices $A^{(pq)}_{ij}$, $B^{(pq)}_{ij}$  are given in terms of the $\Omega$-integrals defined as
\bq
\Omega_{ij}^{(n,r)}=\sqrt{\f{\bk T}{8\pi m_{ij}}}\int_0^\infty Q^{(n)}_{ij}\;\varepsilon^{r+1} e^{-\varepsilon}\md\varepsilon ,
\lae{AY}\ee
where $\varepsilon$ is the dimensionless energy of interacting particles
\bq
\varepsilon=\f{E}{\bk T},\quad E=\f12 m_{ij}|{\bs c}_i-{\bs c}_j|^2,\quad
  i,j=1,2,
\lae{AJ}\ee
$m_{ij}={m_im_j}/(m_i+m_j)$ is the reduced mass of interaction particles. The  transport cross sections $Q^{(n)}_{ij}$ for two different particles are calculated in terms of scattering phase shifts $\delta_l$ \cite{Mee01}
\begin{equation}
Q^{(n)}_{ij}=\f{2\pi \hbar^2}{m_{ij}E}\sum_{l=0}^\infty \sum_{j=0}^{\lfloor(n-1)/2\rfloor} C_{lj}^{(n)}\sin^2\lt(\de_l-\de_{l+n-2j}\rt),
\lae{AT}\end{equation}
where $\hbar$ is the reduced Planck constant. The coefficients $C^{(n)}_{lj}$ are given by Eq.(12) in the previous paper   \cite{Sha118}. For indistinguishable bosons with spin equal to zero,  Eq.(\ref{AT}) should be modified by retaining only even indices $l$ and multiplying the expression (\ref{AT}) by the factor 2.

The numerical scheme to calculate the phase shifts $\delta_l$ is given in details in our previous works \cite{Sha118,Sha126} so that here, only some its improvements will be described. As is known, the the Schr\"odinger equation is solved for relatively small values of the index $l$, i.e. $l\le l_q$. Then, the semi-classical WKB method  \cite{Joa01,Shi07} is used in case of large values of $l$, i.e. $l>l_q$. The value of $l_q$ for the transition from the quantum approach to the semi-classical one  depends on the interaction energy $E$ and species of interacting gases. In the present work, the spherical Bessel and Neumann functions used in the quantum approach have been calculated with the quadruple precision, which allowed us to increase significantly the transition value $l_q$. Now, it is given by the expression
\begin{equation}
l_q=  \lfloor A \cdot E^{1/4}\rfloor,
 \lae{AA}\end{equation}
where the energy $E$ is measured in kelvin. However, its value must be always smaller than $l_{q,max}$. The values of the parameter $A$ and that of the limit $l_{q,max}$ are  given in Table \ref{tabA}.

\begin{table}
\caption{Parameters in Eqs.(\ref{AA})-(\ref{AC}), references about potentials used in the present work, and relative uncertainties $u_{ij}$ of viscosity and thermal conductivity due to potential.}
\begin{tabular}{c |cccc| cc}
  \hline
pair  & \multicolumn{4}{c}{Parameters} & \\
   &$A$(K$^{-1/4}$) & $l_{q,max}$ & $B$(K$^{1/2}$) &  $R_0$ (nm) & Refs. & $u_{ij}$\\ \hline
He-He  & 18 &  450 & 1.3 & 0.264099 &  \cite{Cen01} & $10^{-5}$ \\
Ne-Ne  & 30 &  600 & 2.0 & 0.276125 &  \cite{Hel03} & $10^{-3}$\\
Ar-Ar  & 60 &  1200& 3.0 & 0.335772 &  \cite{Pat02} & $10^{-4}$\\
Kr-Kr  & 70 &  1600& 3.5 & 0.358089 &  \cite{Jag03} & $10^{-3}$\\
He-Ne  & 22 &  600 & 1.4 & 0.269879 &  \cite{Cac01} & $10^{-3}$ \\
He-Ar  & 22 &  600 & 2.0 & 0.311691 &  \cite{Cac01} & $10^{-4}$\\
He-Kr  & 22 &  600 & 2.0 & 0.328702 &  \cite{Jag04} & $10^{-3}$\\
Ne-Ar  & 40 &  900 & 2.3 & 0.312206 &  \cite{Cac01} & $10^{-4}$\\
Ne-Kr  & 50 &  950 & 2.3 & 0.326543 &  \cite{Hal01} & $10^{-3}$\\
Ar-Kr  & 70 &  1400 &3.5 & 0.347762 &  \cite{Hal01} & $10^{-3}$\\ \hline
\end{tabular}\lae{tabA}\end{table}

The quadruple precision also allowed to increase the point $r_m$ where the phase shift is calculated. In the present work, it is given by
\begin{equation}
r_m=10^3\;R_0 \f{B}{\sqrt{E}},
 \lae{AC}\end{equation}
where the energy $E$ is measured in kelvin and $R_0$ is the zero point of the interatomic potential, i.e., $V(R_0)=0$. The values of $B$ and $R_0$ are given in Table \ref{tabA}. The value of $r_m$ does not exceed the limit $10^3R_0$. The integration step is given by
\begin{equation}
\Delta r= R_0 \f{C}{\sqrt[3]{E}},\quad C=10^{-3}\mbox{ K}^{1/3}.
 \lae{AB}\end{equation}
The upper limit for $\Delta r$ is 10$^{-3}R_0$.

The fundamental constants, such as the Bohr radius, atomic mass constant, and Hartree energy are taken from the CODATA-2014 recommended values  \cite{Moh03}. The atomic masses $m_{\mbox{\tiny He}}=4.002 602$, $m_{\mbox{\tiny Ne}}=20.1797$, $m_{\mbox{\tiny Ar}}=39.948$, and $m_{\mbox{\tiny Kr}}=83.798$  measured in the atomic mass constant are taken from Ref. \cite{Mei02}.

The transport cross sections $Q^{(n)}_{ij}(E)$ for the six kinds of collisions have been calculated once for many values of the energy $E$. These quantities are smooth functions of the energy $E$ in the range $E>100$ K, while they have unpredictable behaviours at the small energies. The cross sections $Q_{ii}^{(2)}$ ($i=1,2$) for collisions between the identical particles, namely, Ne-Ne, Ar-Ar,and Kr-Kr are depicted in Figure \ref{figA}, which shows that all of them  have many sharp peaks. The  cross sections $Q_{12}^{(1)}$ for collisions between different particles, {\it viz}., He-Kr, Ne-Kr, and Ar-Kr plotted in Figure \ref{figB} also have many sharp peaks. Such behaviours of the transport cross sections represent a difficulty to calculate the $\Omega$-integrals (\ref{AY}) so that the energy nodes should be distributed non-uniformly. Here, we use a larger number of the nodes $E_m$ than that used previously \cite{Sha118,Sha126}, that is
\begin{equation}
E_m= 2\,(1.00025^m-1), \quad 1\le m\le 48000.
\lae{AP}\end{equation}
Then, the $\Omega$-integrals (\ref{AY}) are calculated  using these knots by the simple trapezoidal rule.

\begin{figure}
\includegraphics{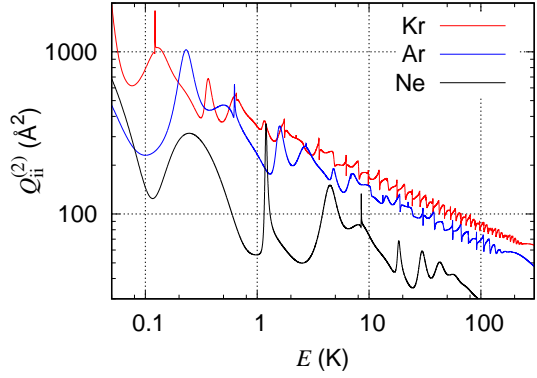}
\caption{Transport cross section $Q_{ii}^{(2)}$ for Ne, Ar, and Kr vs. collision energy $E$.}
\lae{figA}\end{figure}

\begin{figure}
\includegraphics{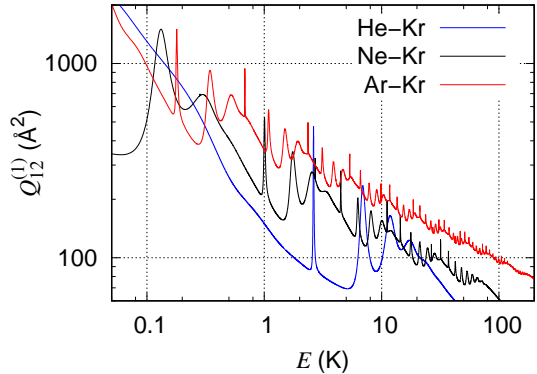}
\caption{Transport cross section $Q_{12}^{(1)}$ for He-Kr, Ne-Kr, and Ar-Kr collisions vs. energy $E$.}
\lae{figB}\end{figure}

\section{Potentials}

According to Ref. \cite{Sha126}, the phase shifts used to calculate the $\Omega$-integrals substituted into the matrices $A_{ij}^{(pq)}$  and  $B_{ij}^{(pq)}$ are obtained from the Schr\"odinger equation containing the interatomic potential. Nowadays, there are many papers reporting {\it ab initio} potentials for homogeneous and heterogeneous dimers of four noble gases, see e.g. Refs. \cite{Cen01,Hel03,Pat02,Jag03,Cac01,Jag04,Hal01,Cyb01,Jag02}. The most reliable of then have been chosen for our calculations. The papers containing potentials used in the present work for main calculations are listed in the sixth column of Table \ref{tabA}.

\section{Uncertainty}

There are two types of uncertainties in the present calculations. The first uncertainty is caused by numerical errors and the second uncertainty is related to interatomic potentials used in the calculations. In this section, these two uncertainties are analyzed and estimated separately.

There are several sources of numerical error in the transport coefficients such as: order of approximation $N$ in Eqs.(\ref{AN1}) and (\ref{AN2}), the value of the parameter $l_q$ for transition from the purely quantum approach to the semi-classical one, the point $r_m$ where the phase shift is calculated, the step of integration $\Delta r$ to solve the  Schr\"odinger equation, the node distribution of the energy (\ref{AP}). The contribution of each error source has been estimated by varying the above mentioned parameters. The main calculations have been carried out for the order approximation $N=10$, the nodes given by (\ref{AP}), the parameters $l_q$, $r_m$, and $\Delta r$ given by Eq.(\ref{AA}), (\ref{AC}), and (\ref{AB}), respectively. Then, test calculations have been carried for $N=12$ with the nodes twice rarer than (\ref{AP}), decreasing $l_q$ and $r_m$ by the factor $0.8$, and increasing $\Delta r$ by the factor 1.5. An analysis of the test results showed that the main numerical contribution into the viscosity comes from the node distribution, which is equal to $\times 10^{-6}$. In case of the thermal conductivity, the main contribution in to the error budget is that because of the approximation order $N$ and equal to $10^{-5}$. In fact, the convergence with respect to the order $N$ for the viscosity is significantly higher than that for the thermal conductivity, especially, in case of the helium-krypton mixture \cite{Jag04}. All other source of the numerical error is orders of magnitude smaller. Thus, the total numerical error can be assumed to be $\times 10^{-6}$ for the viscosity and $10^{-5}$ for the thermal conductivity.

The potentials used in the present work have different degree of their uncertainty. For instance, the helium-helium potential causes the relative uncertainty of $10^{-5}$ in the transport coefficients \cite{Cen01}. At the moment, this is the most exact potential among all other available in the literature. The uncertainties of $\mu$ and $\kappa$ due to the neon-neon \cite{Bic01} and helium-neon \cite{Cac01} potentials estimated in Ref. \cite{Sha118} are equal to $10^{-3}$ over the temperature range considered here. The argon-argon \cite{Pat02}, helium-argon \cite{Cac01}, and neon-argon \cite{Cac01} potentials lead the relative uncertainties in the coefficients $\mu$ and $\kappa$ about $10^{-4}$ according to estimations in Ref. \cite{Sha126}. The uncertainties of the krypton-krypton and helium-krypton potentials estimated in Ref. \cite{Jag04} are equal to  $10^{-3}$. The uncertainties of the neon-krypton and argon-krypton potentials proposed in Ref. \cite{Hal01} were not analyzed previously. However, the krypton-krypton potential elaborated in the same work \cite{Hal01} causes the uncertainty of $10^{-3}$ that can be used as the uncertainty estimation for the  neon-krypton and argon-krypton potentials. All relative uncertainties are summarized in Table \ref{tabA}. As has been mentioned above, the potential is used to calculate the bracket integrals in Eqs.(\ref{AO})-(\ref{AI}). Thus, each potential of interatomic interaction between species $i$ and $j$ contributes into the mixture transport coefficients proportionally to $x_ix_j$. Then, the total relative uncertainty can be estimated by
\begin{equation}
u=\sqrt{\sum_{i=1}^K\sum_{j=i}^K (x_ix_j u_{ij})^2},
\lae{BG}\end{equation}
where the uncertainties $u_{ij}$ are given in Table \ref{tabA}.

\section{Results and discussions}

\subsection{Remarks}

In this section, the numerical results on the viscosity and thermal conductivity are presented and compared with some previously published works. Mainly, the present results will be compared with those reported by Kestin {\it et al.}  \cite{Kes01} who analyzed an extensive database of the transport coefficients of all possible mixtures of noble gases. The authors of Ref. \cite{Kes01} obtained an empirical expressions of the coefficients and estimated their uncertainty by comparing with experimental data published before 1984. During the last decade, new experimental data of viscosity of some single gases were reported \cite{Ber15,Ber14}, but no significant progress has been done in measurements of the thermal conductivity and viscosity of gaseous mixture. Thus, the most of the numerical results obtained here will be compared with those reported by Kestin {\it et al.}  \cite{Kes01}. When possible, a comparison with more accurate theoretical and experimental results is performed. 

\subsection{Binary mixture}

Some binary mixtures, namely, helium-neon, helium-argon, neon-argon were considered in our previous papers \cite{Sha118,Sha126}, where the viscosity and thermal conductivity were calculated with the relative numerical error less than $10^{-5}$ using the quantum approach. The authors of Ref. \cite{Jag04} reported numerical results on the transport coefficients for the helium-krypton  mixture declaring the relative uncertainty about $10^{-3}$. Some binary mixtures were considered also in Ref. \cite{Son49} without an estimation of uncertainty. Below, numerical results on the binary mixtures not considered in our previous papers \cite{Sha118,Sha126}, namely, helium-krypton, neon-krypton, and argon-krypton are presented and compared with those reported in other works  \cite{Kes01,Jag04,Son49}.

The numerical values of viscosity and thermal conductivity of the helium-krypton mixture including pure krypton ($x_1=0$) are reported in Table \ref{tabE}. Since the previous results on pure krypton  \cite{Jag04} are based on the classical theory of interatomic collisions, it is worth to estimate the influence of the quantum effect on the transport coefficients. For this purpose, the transport cross sections $Q_{ij}^{(n)}$ in the $\Omega$-integrals (\ref{AY}) have been calculated applying the classical theory of interatomic collisions \cite{Sha110} for the nodes (\ref{AP}) of the energy $E$ with the relative numerical error less than $10^{-5}$. A comparison of the viscosity based on the quantum approach with that based on the classical one is shown in Figure \ref{figW}. The deviation of the numerical results reported in Ref. \cite{Jag04} from those reported here is also plotted in Figure \ref{figW}. Our results based on the classical approach are in agreement with those reported in Ref.  \cite{Jag04} within $7\times10^{-5}$ that is smaller than the accuracy declared by the authors of Ref. \cite{Jag04}. The plot depicted in Figure \ref{figW} shows that the influence of quantum effects reaches the order $5\times 10^{-4}$, i.e., it exceeds the numerical accuracy of the present work. The measured value of the krypton viscosity at the temperature $t=25^\circ$C reported by Berg \& Burton \cite{Ber14} has the relative uncertainty of  $3\times10^{-4}$ and represents the most exact experimental results till now. The experimental uncertainty of this value plotted by cross in Figure \ref{figW} has the same order as the quantum effect. The deviation of the experimental value from that obtained in the present work is equal to 0.6$\times 10^{-4}$ that represents  the smallest deviation among all theoretical results reported till now.

J\"ager \& Bich  \cite{Jag04} proposed an {\it ab initio} potential for the pair helium-krypton considered by us as the most reliable one. Their numerical data on transport coefficients of the helium-krypton mixture are most exact among all data available in the open literature. They estimated the standard uncertainty of the viscosity to be 0.14\% and that of thermal conductivity values to be 0.2\%. Moreover, they computed all collision integrals for the Kr–Kr atom pair classically, while for Kr–He and He–He collisions, they employed the quantum-mechanical approach. Since we employed the same potentials for He-He, Kr-Kr, and He-Kr collisions as those used by  J\"ager \& Bich  \cite{Jag04}, the use of the classical approach is the main difference of their results from those reported here. Moreover, the present results have been obtained with higher numerical accuracy. The deviations of the results by  J\"ager \& Bich  \cite{Jag04} from the present ones are plotted in Figure \ref{figV}. The deviations are within the uncertainty declared by the authors of Ref. \cite{Jag04}, but they are significantly larger than the numerical error of the present results. The maximum deviation for the viscosity is 0.04\% at $T=$70 K and $x_1 = 0.2$. The greatest deviation of the thermal conductivity is about 0.14\% at 1500 K and $x_1 = 0.2$. The larger deviation of the thermal conductivity is due to its slow convergence with respect to the approximation order $N$ in Eq.(\ref{AN1}).

\begin{table*}
\caption{Viscosity $\mu$ and thermal conductivity $\kappa$  vs. temperature $T$ and molar fraction $x_1$ of helium for He-Kr mixture.}
 \begin{tabular}{r|cccc|cccc} \hline
 &    \multicolumn{4}{c|}{$\mu$($\mu$Pa$\cdot$s)}  &
         \multicolumn{4}{c}{$\kappa$(mW/(m$\cdot$K)) } \\
       $T$(K)& $x_1$=0.& 0.25& 0.5      & 0.75  & $x_1$=0.& 0.25& 0.5      & 0.75   \\ \hline
  50.~~ &	4.95443	&	5.35481	&	5.88556	&	6.52356	&	1.84625	&	5.58588	&	11.4855	&	22.1846  \\
 100.~~ &	8.88984	&	9.56212	&	10.3950	&	11.2229	&	3.30924	&	9.75904	&	19.6828	&	36.9559  \\
 200.~~ &	17.3212	&	18.1263	&	18.9889	&	19.4045	&	6.44504	&	17.0077	&	33.0351	&	60.3896  \\
 300.~~ &	25.4260	&	26.1709	&	26.8203	&	26.5629	&	9.46314	&	23.3654	&	44.4019	&	80.1937  \\
 500.~~ &	39.4057	&	39.9862	&	40.2174	&	38.8167	&	14.6808	&	34.3240	&	64.0522	&	114.661  \\
 800.~~ &	56.4411	&	56.9211	&	56.8114	&	54.3162	&	21.0528	&	48.2624	&	89.4989	&	159.827  \\
1000.~~ &	66.2527	&	66.7421	&	66.5366	&	63.5634	&	24.7243	&	56.5998	&	104.940	&	187.457  \\
2000.~~ &	106.770	&	107.678	&	107.636	&	103.531	&	39.8743	&	92.7663	&	173.135	&	310.695  \\
3000.~~ &	140.504	&	142.049	&	142.581	&	138.231	&	52.4740	&	124.330	&	233.649	&	421.098  \\
5000.~~ &	199.071	&	202.081	&	204.185	&	200.402	&	74.3326	&	181.240	&	344.195	&	624.390  \\ \hline
\end{tabular}\lae{tabE}\end{table*}

\begin{figure}
  \centering
  \includegraphics{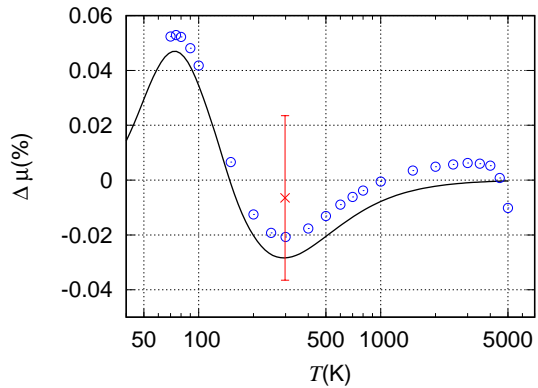}
\caption{Deviation of viscosity $\Delta\mu(\%)=(\mu_{\mbox{\tiny O}}/\mu_{\mbox{\tiny P}}-1)\times100$ of pure krypton based on other theoretical results (subscript ``O'') from that calculated in the present work (subscript ``P"): solid line - $\mu_{\mbox{\tiny O}}$ calculated by classical approach using the same potential as for $\mu_{\mbox{\tiny P}}$; circles - numerical values of  $\mu_{\mbox{\tiny O}}$ reported in Ref. \cite{Jag04}; cross - experimental results on $\mu_{\mbox{\tiny O}}$ reported in Ref. \cite{Ber14}.}
\lae{figW}\end{figure}

\begin{figure*}
\includegraphics{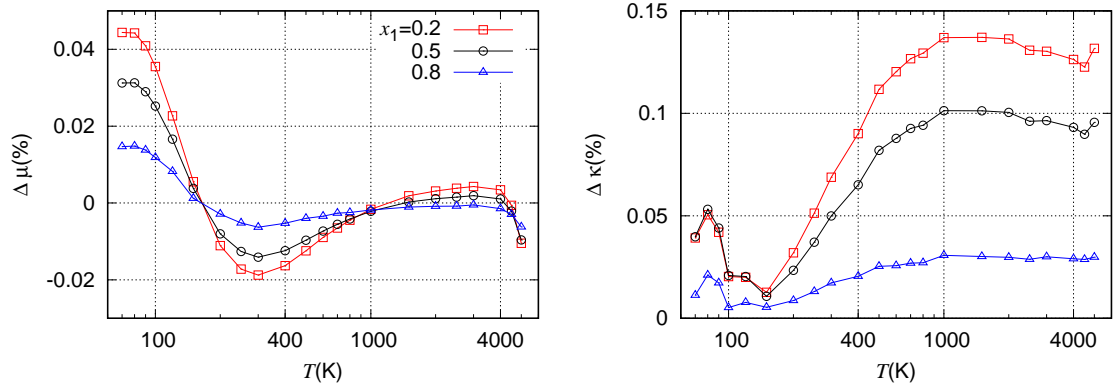}
\caption{Deviation of viscosity (left) and thermal conductivity (right) of helium-krypton mixture reported by J\"ager \& and Bich  \cite{Jag04} (subscript ``J" ) from those calculated in the present work (subscript ``P"), $\Delta C(\%)=(C_{\mbox{\tiny J}}/C_{\mbox{\tiny P}}-1)\times100$, $C=\mu,\kappa$.}
\lae{figV}\end{figure*}

The transport coefficients $\mu$ and $\kappa$ for the neon-krypton mixture are reported in Table \ref{tabI}. In case of pure neon ($x_1=1$), the values of $\mu$ and $\kappa$ are exactly the same as those reported previously \cite{Bic02,Sha118} so that they are omitted in the present work. The most reliable data for this mixture are reported by Kestin {\it et al.}  \cite{Kes01} with the uncertainty being 0.3\% and 0.7\% for   the viscosity and thermal conductivity, respectively. A comparison of these data with the present results is performed in Figure \ref{figC}. As one can notice, the deviations for both viscosity and thermal conductivity are within $\pm$1\% in the temperature range 50 K$\le T\le$2000 K, while they jump up to -3 \% at $T=3000$ K.  Anyhow, the deviations exceed the uncertainties estimated in Ref. \cite{Kes01} and, especially, the total uncertainty of the present results.

\begin{table*}
\caption{Viscosity $\mu$ and thermal conductivity $\kappa$  vs. temperature $T$ and molar fraction $x_1$ of neon for Ne-Kr mixture.}
\begin{tabular}{r|ccc|ccc} \hline
          &    \multicolumn{3}{c|}{$\mu$($\mu$Pa$\cdot$s)}  &
         \multicolumn{3}{c}{$\kappa$(mW/(m$\cdot$K)) } \\
       $T$(K)& $x_1$=0.25& 0.5      & 0.75  & 0.25& 0.5      & 0.75  \\ \hline
50.~~~ & 5.41149 & 6.00505 & 6.79213 & 3.07228 & 4.8161 & 7.48881 \\
100.~~~ & 9.93136 & 11.2223 & 12.7867 & 5.59787 & 8.84097 & 13.7915\\
200.~~~ & 18.8837 & 20.6935 & 22.6437 & 10.2917 & 15.6929 & 23.8155\\
300.~~~ & 27.0732 & 28.9004 & 30.7317 & 14.4565 & 21.4706 & 31.9872\\
500.~~~ & 40.9781 & 42.6395 & 44.1657 & 21.516 & 31.1699 & 45.6578\\
800.~~~ & 57.9745 & 59.5477 & 60.906 & 30.2509 & 43.2962 & 62.8854\\
1000.~~~ & 67.8375 & 69.4528 & 70.825 & 35.3733 & 50.4856 & 73.1658\\
2000.~~~ & 108.994 & 111.275 & 113.233 & 57.0164 & 81.2616 & 117.465\\
3000.~~~ & 143.551 & 146.715 & 149.505 & 75.3856 & 107.652 & 155.616\\
5000.~~~ & 203.834 & 208.848 & 213.412 & 107.669 & 154.310 & 223.176\\ \hline
       \end{tabular}\lae{tabI}\end{table*}

\begin{figure*}
\includegraphics{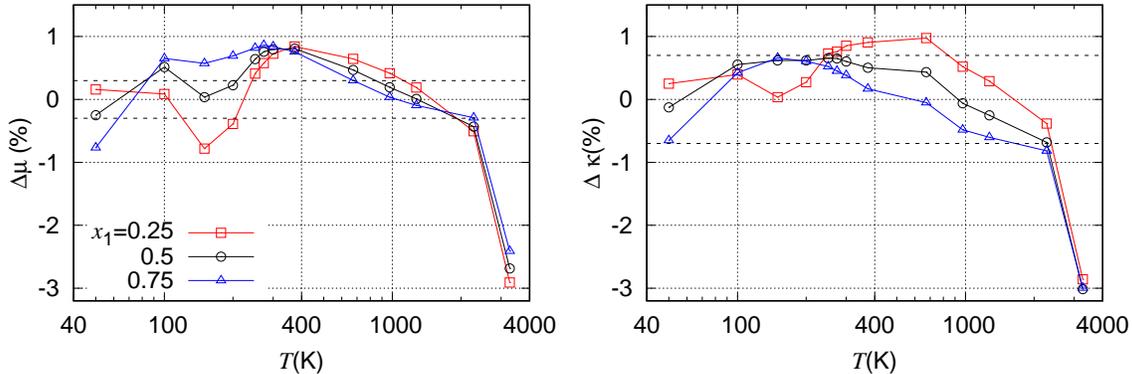}
\caption{Deviation of viscosity (left) and thermal conductivity (right) of neon-krypton mixture
reported by Kestin {\it et al.} \cite{Kes01} (subscript ``K") from those calculated in the present work (subscript ``P"), $\Delta C(\%)=(C_{\mbox{\tiny K}}/C_{\mbox{\tiny P}}-1)\times100$, $C=\mu,\kappa$, dashed lines - uncertainty declared in Ref. \cite{Kes01}.}
\lae{figC}\end{figure*}

The numerical data on the coefficients $\mu$ and $\kappa$ for the argon-krypton mixture are given in Table \ref{tabJ}. The values of $\mu$ and $\kappa$ for pure argon are the same as those reported in the previous work  \cite{Sha126} and not presented here. The paper by Song {\it et al.}  \cite{Son49} also reported the transport coefficients for the argon-krypton mixture, obtained by the Chapman-Enskog method with the first order approximation, i.e., $N=1$ in Eqs.(\ref{AN1}) and (\ref{AN2}). They did not estimated the uncertainty of their results. Kestin {\it et al.}  \cite{Kes01} provided their results on the argon-krypton mixture with the relative uncertainties being  0.4\% and 0.5\% for the viscosity and thermal conductivity, respectively. Figures \ref{figD} presents the comparison between the previously reported data  \cite{Kes01,Son49} and those calculated here. The deviation of viscosity reported in Ref. \cite{Kes01} varies from -3\% to 1\% that is out of the predicted uncertainty over a wide range of the temperature. The data on viscosity provided in Ref. \cite{Son49} are closer to our results and deviate from them  in the range from -1\% to 2\%. The behavior of the thermal conductivity deviation is very similar to that of the viscosity.

\begin{table*}
\caption{Viscosity $\mu$ and thermal conductivity $\kappa$  vs.  temperature $T$ and molar fraction $x_1$ of argon for Ar-Kr mixture}
      \begin{tabular}{r|ccc|ccc} \hline
          &    \multicolumn{3}{c|}{$\mu$($\mu$Pa$\cdot$s)}  &
         \multicolumn{3}{c}{$\kappa$(mW/(m$\cdot$K)) } \\
       $T$(K)& $x_1$=0.25& 0.5      & 0.75  & 0.25& 0.5      & 0.75  \\ \hline
50.~~~ & 4.84845 & 4.71323 & 4.54053 & 2.12785 & 2.4679 & 2.8782\\
100.~~~ & 8.78483 & 8.63446 & 8.42282 & 3.87782 & 4.55498 & 5.36563\\
200.~~~ & 17.141 & 16.8652 & 16.4594 & 7.51387 & 8.81708 & 10.4149\\
300.~~~ & 25.0243 & 24.4701 & 23.7146 & 10.9148 & 12.7115 & 14.9402\\
500.~~~ & 38.4827 & 37.3122 & 35.8266 & 16.7315 & 19.3028 & 22.5157\\
800.~~~ & 54.8465 & 52.897 & 50.5031 & 23.8417 & 27.3529 & 31.7413\\
1000.~~~ & 64.2876 & 61.9068 & 59.0078 & 27.9568 & 32.0245 & 37.1008\\
2000.~~~ & 103.412 & 99.3825 & 94.5187 & 45.053 & 51.5177 & 59.5207\\
3000.~~~ & 136.088 & 130.78 & 124.362 & 59.3561 & 67.8875 & 78.3863\\
5000.~~~ & 192.921 & 185.486 & 176.444 & 84.2641 & 96.4559 & 111.339\\ \hline
       \end{tabular}\lae{tabJ}\end{table*}

\begin{figure*}
\includegraphics{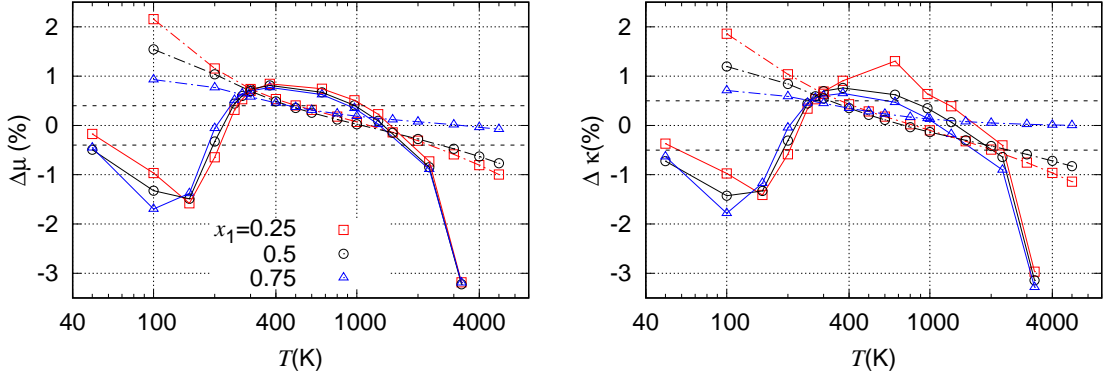}
\caption{Deviation of viscosity (left) and thermal conductivity (right) of argon-krypton mixture reported in other papers  (subscript ``O")  from those calculated in the present work (subscript ``P"), $\Delta C(\%)=(C_{\mbox{\tiny O}}/C_{\mbox{\tiny P}}-1)\times100$, $C=\mu,\kappa$: solid lines - Ref. \cite{Kes01}, point-dashed lines  - Ref. \cite{Son49}, dashed lines - uncertainty declared in Ref. \cite{Kes01}.}
\lae{figD}\end{figure*}

\subsection{Ternary mixtures}

The numerical data on the viscosity and thermal conductivity of ternary mixtures He-Ne-Ar, He-Ne-Kr, He-Ar-Kr, and Ne-Ar-Kr are given in Tables \ref{tabK}, \ref{tabL}, \ref{tabM}, and \ref{tabF}, respectively. First, the equimolar  mixtures ($x_1=x_2=x_3$) is considered, then three situations are reported when one species has a small (0.1) fraction, while two other species have the same fractions equal to 0.45.  Kestin {\it et al.} reported data on the transport coefficients for equimolar ternary mixtures with the relative uncertainty of 0.3\% for the viscosity and 0.7\% for the thermal conductivity. The deviations of their data from those presented here are plotted in Figure \ref{figG}.  The deviations of the all viscosities have a similar behavior for temperatures above 200 K. Mostly, the deviations slightly exceed the value of 0.3\% declared in Ref. \cite{Kes01}. However, they are significant, i.e., about  3.4\%, at $T=3273$ K. In case of the thermal conductivity, the behaviors of all deviations are also similar to each other, except that of the neon-argon-krypton mixture. The deviation of this mixture is within the value of 0.7\% estimated in Ref. \cite{Kes01}. The deviations of all other mixtures are significantly larger than 0.7\% in the temperature range $T>$ 800 K, and reach the magnitude almost 6\%.

\begin{table*}
\caption{Viscosity $\mu$ and thermal conductivity $\kappa$  vs.  temperature $T$ and molar fractions $x_1$ of He and $x_2$ of Ne for ternary mixture of He-Ne-Ar.}
      \begin{tabular}{r|cccc|cccc}  \hline
          &    \multicolumn{4}{c|}{$\mu$($\mu$Pa$\cdot$s)}  &
         \multicolumn{4}{c}{$\kappa$(mW/(m$\cdot$K)) } \\
& $x_1=1/3$& 0.1   & 0.45 & 0.45 &  1/3     & 0.1      & 0.45 & 0.45 \\
$T$(K)& $x_2=1/3$& 0.45  & 0.1  & 0.45 & 1/3      & 0.45   & 0.1 & 0.45  \\ \hline
50.~~~ & 6.10706 & 5.77222 & 5.54972 & 7.16617 & 13.2148 & 8.15544 & 13.802 & 19.9698\\
100.~~~ & 11.1674 & 10.8747 & 10.0296 & 12.6352 & 23.0374 & 14.9100 & 23.792 & 33.7948\\
200.~~~ & 19.5115 & 19.5211 & 17.839 & 21.0243 & 38.6649 & 25.9951 & 39.9129 & 55.2288\\
300.~~~ & 26.3577 & 26.6133 & 24.3983 & 27.8514 & 51.6279 & 35.1051 & 53.4178 & 73.0776\\
500.~~~ & 37.7725 & 38.3243 & 35.3435 & 39.4203 & 73.773 & 50.3568 & 76.5871 & 103.916\\
800.~~~ & 52.1081 & 52.8657 & 49.0155 & 54.2326 & 102.308 & 69.5952 & 106.525 & 144.100\\
1000.~~~ & 60.6531 & 61.4741 & 57.1359 & 63.1596 & 119.586 & 81.094 & 124.685 & 168.585\\
2000.~~~ & 97.4822 & 98.2767 & 92.0357 & 102.093 & 195.563 & 130.833 & 204.798 & 277.052\\
3000.~~~ & 129.25 & 129.769 & 122.106 & 136.034 & 262.513 & 173.908 & 275.695 & 373.317\\
5000.~~~ & 185.651 & 185.298 & 175.52 & 196.785 & 383.744 & 250.673 & 404.673 & 548.69\\ \hline
\end{tabular}\lae{tabK}\end{table*}

\begin{table*}
\caption{Viscosity $\mu$ and thermal conductivity $\kappa$  vs.  temperature $T$ and molar fractions $x_1$ of He and $x_2$ of Ne for ternary mixture He-Ne-Kr.}
      \begin{tabular}{r|cccc|cccc}  \hline
          &    \multicolumn{4}{c|}{$\mu$($\mu$Pa$\cdot$s)}  &
         \multicolumn{4}{c}{$\kappa$(mW/(m$\cdot$K)) }  \\
& $x_1=1/3$& 0.1   & 0.45 & 0.45 &  1/3     & 0.1      & 0.45 & 0.45 \\
$T$(K)& $x_2=1/3$& 0.45  & 0.1  & 0.45 & 1/3      & 0.45   & 0.1 & 0.45  \\ \hline
50.~~~ & 6.47799 & 6.13932 & 6.05689 & 7.35753 & 11.2683 & 6.44034 & 11.3982 & 18.8917\\
100.~~~ & 11.775 & 11.3923 & 10.8031 & 13.0458 & 19.5261 & 11.5590 & 19.5821 & 31.9538\\
200.~~~ & 20.9794 & 20.8139 & 19.5947 & 21.945 & 32.7818 & 20.0589 & 32.8666 & 52.3048\\
300.~~~ & 28.8565 & 28.9485 & 27.4485 & 29.2396 & 43.8987 & 27.2039 & 44.1296 & 69.2914\\
500.~~~ & 42.1115 & 42.5821 & 40.807 & 41.5787 & 62.9779 & 39.3003 & 63.5616 & 98.6546\\
800.~~~ & 58.6333 & 59.4147 & 57.3788 & 57.3115 & 87.5978 & 54.618 & 88.7051 & 136.920\\
1000.~~~ & 68.404 & 69.2996 & 67.1177 & 66.7682 & 102.511 & 63.7804 & 103.957 & 160.24\\
2000.~~~ & 110.136 & 111.16 & 108.415 & 107.901 & 168.15 & 103.46 & 171.27 & 263.576\\
3000.~~~ & 145.884 & 146.731 & 143.616 & 143.682 & 226.096 & 137.904 & 230.931 & 355.35\\
5000.~~~ & 209.100 & 209.229 & 205.739 & 207.643 & 331.286 & 199.484 & 339.748 & 522.694\\ \hline
\end{tabular}\lae{tabL}\end{table*}

\begin{table*}
\caption{Viscosity $\mu$ and thermal conductivity $\kappa$  vs.  temperature $T$ and molar fractions $x_1$ of He and $x_2$ of Ar for ternary mixture of He-Ar-Kr.}
      \begin{tabular}{r|cccc|cccc}  \hline
          &    \multicolumn{4}{c|}{$\mu$($\mu$Pa$\cdot$s)}  &
         \multicolumn{4}{c}{$\kappa$(mW/(m$\cdot$K)) }  \\
& $x_1=1/3$& 0.1   & 0.45 & 0.45 &  1/3     & 0.1      & 0.45 & 0.45 \\
$T$(K)& $x_2=1/3$& 0.45  & 0.1  & 0.45 & 1/3      & 0.45   & 0.1 & 0.45  \\ \hline
50.~~~ & 5.29299 & 4.86662 & 5.69931 & 5.34308 & 8.20499 & 3.88315 & 10.4219 & 11.9922\\
100.~~~ & 9.55002 & 8.88415 & 10.131 & 9.61958 & 14.3276 & 6.9926 & 17.9535 & 20.7109\\
200.~~~ & 17.8363 & 17.1462 & 18.6296 & 17.5094 & 24.6109 & 12.7782 & 30.3021 & 34.9706\\
300.~~~ & 25.275 & 24.7204 & 26.3361 & 24.3398 & 33.4014 & 17.9039 & 40.8225 & 47.0045\\
500.~~~ & 37.8235 & 37.5033 & 39.461 & 35.773 & 48.4745 & 26.6186 & 58.9726 & 67.6638\\
800.~~~ & 53.2406 & 53.0623 & 55.6774 & 49.9386 & 67.7704 & 37.4801 & 82.4103 & 94.3094\\
1000.~~~ & 62.2519 & 62.0823 & 65.1739 & 58.2948 & 79.3913 & 43.8888 & 96.606 & 110.45\\
2000.~~~ & 100.248 & 99.7312 & 105.282 & 93.9376 & 130.234 & 71.2189 & 159.152 & 181.562\\
3000.~~~ & 132.471 & 131.372 & 139.361 & 124.469 & 174.923 & 94.6643 & 214.517 & 244.457\\
5000.~~~ & 189.118 & 186.623 & 199.387 & 178.509 & 255.866 & 136.305 & 315.432 & 358.896\\ \hline
\end{tabular}\lae{tabM}\end{table*}

\begin{table*}
\caption{Viscosity $\mu$ and thermal conductivity $\kappa$  vs.  temperature $T$ and molar fractions $x_1$ of Ne and $x_2$ of Ar for ternary mixture Ne-Ar-Kr.}
      \begin{tabular}{r|cccc|cccc}  \hline
          &    \multicolumn{4}{c|}{$\mu$($\mu$Pa$\cdot$s)}  &
         \multicolumn{4}{c}{$\kappa$(mW/(m$\cdot$K)) }  \\
& $x_1=1/3$& 0.1   & 0.45 & 0.45 &  1/3     & 0.1      & 0.45 & 0.45 \\
$T$(K)& $x_2=1/3$& 0.45  & 0.1  & 0.45 & 1/3      & 0.45   & 0.1 & 0.45  \\ \hline
50.~~~ & 5.41134 & 4.90012 & 5.82345 & 5.57393 & 4.28900 & 2.94238 & 4.65229 & 5.68228\\
100.~~~ & 10.1395 & 9.04599 & 10.8919 & 10.5746 & 7.97242 & 5.44821 & 8.57612 & 10.6585\\
200.~~~ & 19.0126 & 17.4665 & 20.181 & 19.4024 & 14.4974 & 10.3102 & 15.323 & 19.1343\\
300.~~~ & 26.7318 & 25.1099 & 28.2388 & 26.7755 & 20.0515 & 14.6421 & 21.0252 & 26.1505\\
500.~~~ & 39.592 & 37.9595 & 41.7087 & 38.9421 & 29.3379 & 21.9381 & 30.5877 & 37.7925\\
800.~~~ & 55.3398 & 53.5873 & 58.2618 & 53.932 & 40.8567 & 30.8938 & 42.5177 & 52.269\\
1000.~~~ & 64.5434 & 62.6495 & 67.9523 & 62.7549 & 47.6499 & 36.1204 & 49.5805 & 60.8381\\
2000.~~~ & 103.327 & 100.486 & 108.846 & 100.236 & 76.5608 & 58.0883 & 79.7663 & 97.4605\\
3000.~~~ & 136.148 & 132.28 & 143.487 & 132.144 & 101.231 & 76.6477 & 105.616 & 128.796\\
5000.~~~ & 193.645 & 187.765 & 204.207 & 188.211 & 144.703 & 109.159 & 151.275 & 184.059\\ \hline
\end{tabular}\lae{tabF}\end{table*}

\begin{figure*}
\includegraphics{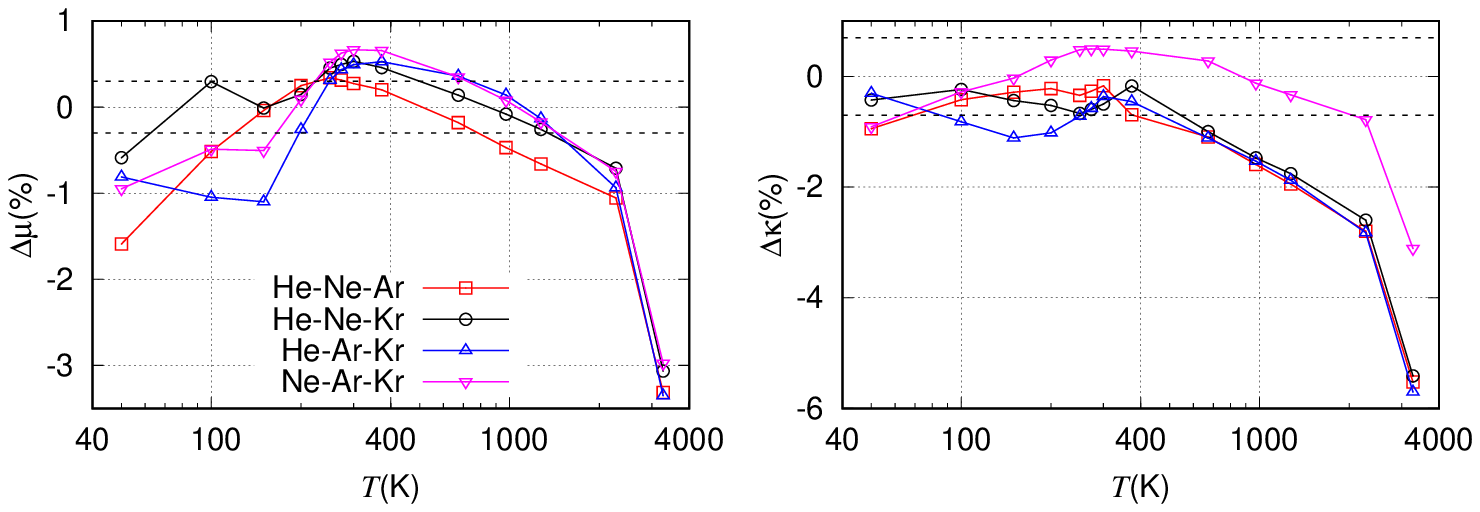}
\caption{Deviations of viscosities (left) and thermal conductivities (right) of equimolar ternary mixtures reported by Kestin {\it et al.}  \cite{Kes01} (subscript ``K") from those calculated in the present work (subscript ``P"), $\Delta C(\%)=(C_{\mbox{\tiny K}}/C_{\mbox{\tiny P}}-1)\times100$, $C=\mu,\kappa$, dashed lines - uncertainty declared in Ref. \cite{Kes01}.}
\lae{figG}\end{figure*}

\subsection{Quaternary mixture}

Numerical data on viscosity and thermal conductivity of the quaternary He-Ne-Ar-Kr mixture are given in Table \ref{tabH} for equimolar composition and for four compositions when the molar fraction of one species is 0.1, while that for the rest of species is 0.3. Kestin {\it et al.}  \cite{Kes01} reported the transport coefficients only for the equimolar mixture with the uncertainties of 0.3\% and 0.7\% for viscosity and thermal conductivity, respectively. Figure \ref{figI} shows the deviation of the data by Kestin {\it et al.} \cite{Kes01} from the present results. In case of viscosity, the deviation is within the uncertainty in the temperature range from 100 K to 1273 K, but it reaches 3.2\% at $T=$3273 K. The deviation of the thermal conductivity is larger than that of the viscosity. At high temperatures, the deviation magnitude reaches  the value of 5.4\%.

\begin{figure}
\includegraphics{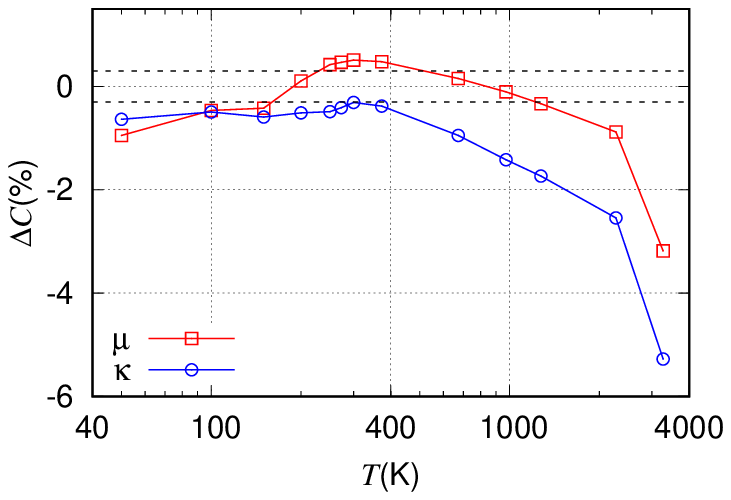}
\caption{Deviation of viscosity and thermal conductivity of equimolar helium-neon-argon-krypton mixture reported by Kestin {\it et al.} (subscript ``K") from those calculated in the present work (subscript ``P"), $\Delta C(\%)=(C_{\mbox{\tiny K}}/C_{\mbox{\tiny P}}-1)\times100$, $C=\mu,\kappa$, dashed lines - uncertainty of thermal conductivity declared in Ref. \cite{Kes01}.}
\lae{figI}\end{figure}

\begin{table*}
\caption{Viscosity $\mu$ and thermal conductivity $\kappa$  vs.  temperature $T$ and molar fractions of helium $x_1$, neon $x_2$, and argon $x_3$ for quaternary  mixture He-Ne-Ar-Kr.}
      \begin{tabular}{r|ccccc|ccccc}  \hline
       & \multicolumn{5}{c|}{$\mu$($\mu$Pa$\cdot$s)} & \multicolumn{5}{c}{$\kappa$(mW/(m$\cdot$K))} \\
       & $x_1=0.25$ & 0.1 & 0.3 & 0.3 & 0.3 & $x_1=0.25$ &  0.1 & 0.3 & 0.3 & 0.3   \\
       & $x_2=0.25$ & 0.3 & 0.1 & 0.3 & 0.3 & $x_2=0.25$ &  0.3 & 0.1 & 0.3 & 0.3   \\
$T$(K) & $x_3=0.25$ & 0.3 & 0.3 & 0.1 & 0.3 & $x_3=0.25$ &  0.3 & 0.3 & 0.1 & 0.3   \\  \hline
50.~~~   & 5.79497 &    5.55685&    5.48530 &    6.18955 &    5.99105 & 8.72551&   5.87628 &  8.39333 & 10.1887 &   11.1836 \\
100.~~~  & 10.6494 &   10.3394 &    9.97736 &   11.3028  &   11.0007  & 15.3859&   10.6418 &  14.7101 & 17.7800 &   19.5962 \\
200.~~~  & 19.4088 &   19.1822 &   18.4563  &   20.3206  &   19.5907  & 26.3442&   18.7734 &  25.2385 & 30.0639 &   33.1505 \\
300.~~~  & 26.9424 &   26.8409 &   25.9352  &   28.0515  &   26.7700  & 35.5593&   25.6491 &  34.1805 & 40.3717 &   44.4457 \\
500.~~~  & 39.5392 &   39.6151 &   38.5018  &   41.0259  &   38.7355  & 51.2660&   37.2498 &  49.4763 & 58.0165 &   63.7190 \\
800.~~~  & 55.1166 &   55.3130 &   53.9782  &   57.1462  &   53.6535  & 71.3473&   51.8548 &  69.0484 & 80.7088 &   88.4801 \\
1000.~~~ & 64.2867 &   64.5117 &   63.0499  &   66.6619  &   62.4984  & 83.4399&   60.5585 &  80.8361 & 94.4249 &  103.443 \\
2000.~~~ & 103.265 &  103.397  &  101.427   &  107.226   &  100.401   & 136.290&   98.1008 & 132.394  &154.640  &  169.100 \\
3000.~~~ & 136.512 &  136.399  &  134.052   &  141.914   &  132.945   & 182.624&   130.580 & 177.672  &207.661  &  226.855 \\
5000.~~~ & 195.118 &  194.340  &  191.471   &  203.176   &  190.551   & 266.239&   188.514 & 259.574  &303.697  &  331.317\\
\hline
\end{tabular}\lae{tabH}\end{table*}

\section{Conclusions}

The viscosity and thermal conductivity of multi-component mixtures composed from helium, neon, argon, and krypton in the limit of low density have been calculated on the basis of {\it ab initio} potentials over the temperature range from 50 K - 5000 K. The Chapman-Enskog method with the 10th order of approximation has been employed. The relative numerical error does not exceed the value $10^{-6}$ and $10^{-5}$ for the viscosity and thermal conductivity, respectively. However, the relative uncertainty of these coefficients related to the potentials reaches $10^{-3}$. It has been shown that the quantum effects in the interatomic collisions of krypton pair affects the transport coefficients within 0.05\% that is about the experimental error reported in Ref. \cite{Ber14}. That is why, all calculations have been carried out using the quantum approach to interatomic collisions. Moreover, the present results on pure krypton are closest to the experimental value \cite{Ber14} among all previous theoretical works. The viscosity and thermal conductivity of binary, ternary and quaternary mixtures have been compared with other theoretical works showing that the present results are most accurate at the moment. The results data in the present work together with those published previously \cite{Cen01,Sha118,Sha126,Bic02} represent the complete database of the viscosity and thermal conductivity of all possible mixtures composed from helium, neon, argon, and krypton over wide ranges of the temperature and mole fractions. 

\section*{Acknowledgments:}

One of the authors (F.S.) acknowledges the Brazilian Agency CNPq for the support of his research, Grant No. 304831/2018-2.

\appendix

\section{Derivation of thermal conductivity expression}

To derive the expression (\ref{QR}) of the steady state thermal conductivity of gaseous mixture composed from $K$ species, we depart from Eq.(6.3-50) of the book by Ferziger and Kaper  \cite{Fer02}
\begin{equation}
\kappa=\f13\bk\sum_{i,j=1}^K x_ix_j\int \wt{{\bs A}}_i \cdot I_{ij}(\wt{{\bs A}}) \md^3c_i,
\lae{DU}\end{equation}
\bq
\wt{{\bs A}}_i={\bs A}_i-\sum_{l=1}^K k_{Tl}{\bs D}_i^l,
\lae{DU1}\ee
where $I_{ij}$ is the linearized collision integral between species $i$ and $j$  \cite{Fer02}, $k_{Tl}$ is the thermal diffusion ratio of species $l$ coupled as
\begin{equation}
\sum_{l=1}^K k_{Tl}=0.
\lae{BY}\end{equation}
The vectors ${\bs A}_i$ and ${\bs D}_i^l$ obey the following Boltzmann equations, see Eqs. (6.3-18) and (6.3-19) from the book \cite{Fer02},
\begin{equation}
\sum_{j=1}^K x_ix_j I_{ij}({\bs A})=\f{1}{n}f^M_i\lt(\C_i^2-\f52\rt){\bs C}_i,
\lae{EC}\end{equation}
\begin{equation}
\sum_{j=1}^K x_ix_j I_{ij}({\bs D}^l)=\f{1}{n_i}f^M_i\lt(\delta_{il}-\f{\rho_i}{\rho}\rt){\bs C}_i,
\lae{DV}\end{equation}
\begin{equation}
f_i^M=n_i\lt(\f{m}{2\pi \bk T}\rt)^{3/2}e^{-\C^2_i},
\lae{AZ}\end{equation}
\bq
 \rho_i=m_in_i,\q \rho=\sum_{i=1}^K\rho_i.
\ee
Combining (\ref{BY}) and (\ref{DV}), we obtain
\begin{equation}
\sum_{j=1}^K x_ix_j I_{ij}\lt(\sum_{l=l}^K k_{Tl}{\bs D}^l\rt)=
\f{k_{Ti}}{n_i}f^M_i{\bs C}_i.
\lae{FD}\end{equation}
A summation of (\ref{EC}) and (\ref{FD}) leads to the integral equation for $\wt{\bs A}$
\begin{equation}
\sum_{j=1}^K x_ix_j I_{ij}(\wt{\bs A})=\f{1}{n}f^M_i\lt(\C_i^2-\f52-\f{k_{Ti}}{x_i}\rt){\bs C}_i.
\lae{DY}\end{equation}
To solve this equation, the variational principle formulated in Ref. \cite{Fer02} is used. First, a trial function ${\bs a}_i({\bs C}_i)$ of the order $N$ is introduced as
\begin{equation}
{\bs a}_i({\bs C}_i)=- \f{15\sqrt{m_i}}{4} \sum_{p=1}^N {a}_i^{(p)}S_{3/2,i}^{(p)}{\bs C}_i,
\lae{FE}\end{equation}
where $S_{3/2,i}^{(p)}$ are given by (\ref{AL}). The terms for $p=0$ are omitted because each species is at rest. To find the coefficients $a_i^{(p)}$, we need to maximize the functional $g$
\begin{equation}
g=\f{8}{225\bk T}\sum_{i,j=1}^K x_ix_j\int {{\bs a}}_i \cdot I_{ij}({{\bs a}}) \md^3c_i
\lae{FF}\end{equation}
under the following constrains
\begin{equation}
\sum_{j=1}^K x_j \int {{\bs a}}_i \cdot I_{ij}({{\bs a}}) \md^3c_i=
\sum_{j=1}^K x_j \int {{\bs a}}_i \cdot I_{ij}(\wt{{\bs A}}) \md^3c_i.
\lae{FK}\end{equation}
These two conditions guarantee that ${\bs a}_i$ tends to $\wt{\bs A}$ in the limit $N\to\infty$. Substituting  (\ref{DY}) and (\ref{FE}) into (\ref{FF}) and (\ref{FK}), we obtain
\begin{equation}
g=\sum_{i,j=1}^K\sum_{p,q=1}^N A_{ij}^{(pq)}a_i^{(p)}a_j^{(q)},
\lae{BA}\end{equation}
and
\begin{equation}
\sum_{j=1}^K \sum_{p,q=1}^N A^{(pq)}_{ij} a_i^{(p)}a_j^{(q)}=\f{x_i}{\sqrt{m_i}}a_i^{(1)},
\lae{FN}\end{equation}
respectively. Using the Lagrangian multipliers to combine Eq.(\ref{FN}) with the conditions $\delta\{g\}=0$ where $g$ is given by (\ref{BA}), the system of algebraic equations (\ref{AN2}) is derived. A substitution of (\ref{FN}) into (\ref{BA}) results the simple expression of $g$
\begin{equation}
g=\sum_{i=1}^K  \f{x_i}{\sqrt{m_i}}a_i^{(1)}.
\lae{BC}\end{equation}
To obtain the thermal conductivity expression (\ref{QR}), the quantity $\wt{\bs A}$ in Eq.(\ref{DU}) is replaced by ${\bs a}_i$ in the form (\ref{FE}) and the obtained expression is compared with (\ref{FF}). Then we see
\begin{equation}
\kappa=\f{75}{8}\bk^2T g.
\lae{BD}\end{equation}
A substitution of (\ref{BC}) into (\ref{BD}) leads to (\ref{QR}).

\end{document}